\documentclass[preprint,aps,amsmath,amssymb,showpacs]{revtex4-1}
\usepackage{graphicx}
\usepackage{comment}

\begin{document}
\title{Monte Carlo simulations of polymers with nearest- and next nearest-neighbor interactions on square and cubic lattices}
\author{Nathann T. Rodrigues}
%\email{nathan.rodrigues@ufv.br}
\author{Tiago J. Oliveira}
\email{tiago@ufv.br}
\affiliation{Departamento de F\'isica, Universidade Federal de Vi\c cosa, 36570-900, Vi\c cosa, MG, Brazil}
\date{\today}

\begin{abstract}
We study a generalized interacting self-avoiding walk (ISAW) model with nearest- and next nearest-neighbor (NN and NNN) interactions on the square and cubic lattices. In both dimensions, the phase diagrams show coil and globule phases separated by continuous transition lines. Along these lines, we calculate the metric $\nu_t$, crossover $\phi_t$ and entropic $\gamma_t$ exponents, all of them in good agreement with the exact values of the $\Theta$ universality class. Therefore, the introduction of NNN interactions does not change the class of the ISAW model, which still exists even for repulsive forces. The growth parameters $\mu_t$ are shown to change monotonically with temperature along the $\Theta$-lines. In the square lattice, the $\Theta$-line has an almost linear behavior, which was not found in the cubic one. Although the region of repulsive NNN interactions, with attractive NN ones, leads to stiff polymers, no evidence of a transition to a crystalline phase was found. 

\end{abstract}

%\pacs{05.40.Fb,05.70.Fh,61.41.+e}

\maketitle

\section{Introduction}
\label{intro}

Polymers in dilute solutions are known to present non-trivial thermodynamic properties \cite{Cloizeaux}. Even in the simplest case of a single flexible homopolymer chain, conformational phase transitions are present. Among then, the most studied is the coil-globule, or collapse, transition: At high temperatures the excluded volume interaction dominates the system, which stays in a swollen coil conformation. As temperature is lowered, monomer-monomer attractive interactions prevail, giving rise to a collapsed (globule) configuration \cite{Flory1,Flory2}. These effects cancel each other at the so-called $\Theta$-point and a continuous coil-globule transition takes place at temperature $T_{\Theta}$, which is known to be a tricritical point \cite{DeGennes1,DeGennes2}. 

The collapse transition has been modeled by interacting self-avoiding walks (ISAWs), where, beyond the excluded volume, a short-range attractive force between non-consecutive monomers in the chain is considered \cite{Carlo,Flory2,DeGennes2}. On the lattice, the standard ISAW model consists in assign an attractive interaction between non-bounded nearest-neighbor (NN) monomers in the walk. This model has been extensively studied by different approaches (see Tables I and II of Refs. \cite{Lee1} and \cite{Lee2} for estimates of $T_{\Theta}$ in 3D and 2D, respectively). The radius of gyration (or, equivalently, the end-to-end distance), $R_N$, of a chain with $N$ monomers is expected to scale near the criticality as \cite{DeGennes1,DeGennes2}
\begin{equation}
 \left\langle R_N^2 \right\rangle \sim N^{2 \nu_t} f\left( \tau N^{\phi_t} \right),
\label{eqScaling}
\end{equation}
where $\nu_t$ and $\phi_t$ are the tricritical metric and crossover exponents, respectively, $\tau \equiv |T - T_{\Theta}|/T_{\Theta}$ and $f(x)$ is a scaling function, which behaves as \cite{meirovitch,meirovitch2,Tesi}
\begin{equation}
f(x) \sim \left \{
\begin{array}{ll}
x^{(2 \nu_{SAW} -2\nu_t)/\phi_t}, & \text{if} \quad x \rightarrow \infty, \\
const., & \text{if} \quad x=0, \\
|x|^{(2/d-2\nu_t)/\phi_t}, & \text{if} \quad x \rightarrow -\infty.
\end{array}
\right.
\label{eqScaling2}
\end{equation}
Therefore, in general, $\left\langle R_N^2 \right\rangle \sim N^{2 \nu}$, with exponents $\nu=\nu_{SAW}$ in the coil phase, $\nu=\nu_t$ at the $\Theta$-point and $\nu=1/d$ in the globule phase. The $\nu_{SAW}$ exponent assume the exact Flory value $\nu_{SAW}=3/(d+2)$ \cite{Flory2} in two dimensions ($d=2$), but deviate from this in $d=3$, where it is $\nu \simeq 0.588$ (for a recent reference see \cite{Clisby}). The $\Theta$-point critical exponents have the exact values $\nu_{t}=4/7$ and $\phi_t=3/7$ in $d=2$ \cite{Carlo,DupSaleur}. In three dimensions, which is the upper critical dimension, the exponents assume the mean-field values $\nu_{t}=\phi_t=1/2$ \cite{Flory2,DeGennes2} and logarithmic corrections to scaling are expected \cite{DeGennes3}.

\begin{comment}
Since the globular structures are dense but disordered (liquid-like), a transition towards an ordered (solid-like) phase could be expected for $T<T_{\Theta}$. In fact, a first-order freezing (disorder-to-order) transition has been observed in some simulation works in three dimensions \cite{Zhou1,Alves,Rampf1,Rampf2,Parsons,Paul,Vogel,Lee1}. A quite similar ordered crystalline phase is also found in semiflexible (stiff) polymers, which may be modeled associating an energy penalty to elementary bends of the chain. Thus, this bending energy is minimized in a crystal (solid-like) phase composed mainly by compact linear stripes, which appears for small temperatures and large enough bending energies (see \cite{Zhou2} and references therein).
\end{comment}

Beyond that classical ISAW model, other approaches have been employed to model the polymer collapse transition. For example, collapsing walks can be obtained by constraining the walk atmospheric statistics, which restrict its degrees of freedom (see \cite{Alvarez} and references therein). Another example is the multiple monomer per site (MMS) model by Krawczyk et al. \cite{Krawczyk}, where lattice sites can be occupied by up to $K$ monomers. For $K=3$, Monte Carlo simulations provided a rich (canonical) phase diagram for this model with coil and globule phases separated either by a $\Theta$-line (a continuous line of $\Theta$-points) or by a line of discontinuous (first-order) transition, both matching at a multicritical point \cite{Krawczyk}. Although it is well accepted that the nature of coil-globule transition changes in this model, the existence of a discontinuous transition had been questioned in \cite{Pablo,Tiago}.

A generalized ISAW model has also been used to model semiflexible polymers, where a repulsive bending force, associated to a bending energy $\epsilon_b<0$, is introduced in the ISAW model (see \cite{Zhou2} and references therein). Beyond the coil and globule phases (separated by a $\Theta$-line), this model presents a crystalline phase and discontinuous globule-crystal (at small $\epsilon_b$) and coil-crystal (at large $\epsilon_b$) transitions are also present in its phase diagram \cite{Zhou2}.

More recently, another generalization of the ISAW model was proposed by Lee et al. \cite{Lee3,Lee4}, which considered attractive interactions, with energies $\epsilon_1$ and $\epsilon_2$, between NN and next-nearest-neighbor (NNN) monomers, respectively. They calculate exactly the partition function zeros for chains with up to $38$ monomers on the square lattice and showed that the crossover exponent is not affected by the NNN interaction. Therefore, it was suggested that the universality class of this system is the same of the simple ISAW model ($\epsilon_2 = 0$) - hereafter called the $\Theta$ universality class.

Since in real polymers the interactions could be larger than one lattice parameter (typical monomer-monomer distances) \cite{Flory2,DeGennes2}, it is very important to determine how robust is the $\Theta$ universality class to changes in the range (by including NNN ones) and nature of interactions. In order to do this, here we study the model of Lee et al. \cite{Lee3} on the square and cubic lattices via Monte Carlo simulations. Growing equilibrium chains with the pruned-enriched Rosenbluth method (PERM) \cite{Grassberger1}, polymers with up to 5000 monomers are studied in a broad range of energies, including negative (repulsive interaction) ones. The canonical phase diagrams (both in $d=2$ and $d=3$) display coil and globule phases separated by lines of $\Theta$-points, even in repulsive interaction regions. Thus, the collapsed phase exists even for negative NN (NNN) energies with large enough positive NNN (NN) ones. In the square lattice, the $\Theta$-line has an almost linear behavior in $\epsilon_1/k_B T \times \epsilon_2/k_B T$ plane, similarly to the findings of Lee et al. \cite{Lee4}.

The rest of this work is organized as follows. In Sec. \ref{defmod} we define the model and the simulation algorithm. The thermodynamic properties of the model on the square and cubic lattices are presented in Sec. \ref{results}. In Sec. \ref{conclusions} our final discussions and conclusions are summarized.

\section{The model and Monte Carlo method}
\label{defmod}

We consider a generalized interacting self-avoiding walk model on square and cubic lattices, where an energy $\epsilon_1$ is assigned to each pair of non-bounded nearest-neighbor (NN) monomers and an energy $\epsilon_2$ is associated to each pair of next-nearest-neighbor (NNN) monomers. Thus, for a given configuration $S=\left\lbrace s_1,s_2,\ldots,s_N \right\rbrace $ of a chain with $N$ monomers, we have the total energy
\begin{equation}
E(S)=-\epsilon_1 M_{1}(S) - \epsilon_2 M_{2}(S),
\label{eqENERGY}
\end{equation}
where $M_{1}$ and $M_{2}$ gives the number of pair of NN and NNN monomers in configuration $S$, respectively. Since the number of conformations of size $N$, $\Omega_N$, increases exponentially with $N$ \cite{Gutt1,Gutt2}, it is very difficult to calculate exactly the canonical partition function $Z_N = \sum_S e^{-E(S)/k_B T}$ for large polymers. Therefore, we will estimate $Z_N$ employing Monte Carlo (MC) simulations.

There are several MC methods to efficiently sample equilibrium configurations of lattice polymers (see \cite{Rensburg} for a review). Here, we will use an improvement of the classical Rosenbluth method \cite{Rosenbluth}, namely, the pruned-enriched Rosenbluth method (PERM) \cite{Grassberger1}. This approach has been employed in several systems and is particularly useful in the study of collapsing polymers \cite{Grassberger2}. 

In the Rosenbluth method, a chain is grown by trying to insert new monomers at \textit{empty} NN sites of its end. This growth follows until the length $N$ is attained or the chain becomes trapped, i. e., all the NN sites of its end are already occupied. Since only empty sites are tried, there is a bias in the chain, which should be corrected by associating a weight to the generated configuration. Considering a thermal (interacting) chain with $n-1$ monomers, if $l_n$ is the number of empty NN sites of the monomer $n-1$, the $k^{th}$ of these sites could be chosen with probability $p_k = e^{-E_{k}/k_B T}/w_{n}^{(k)}$, where $w_n^{(k)} = \sum_{i=1}^{l_n} e^{-E_i/k_B T}$. Thus, the Rosenbluth weight of a configuration $S$ with $N$ monomers is given by
\begin{equation}
 W_{N}(S) = \prod_{n=1}^{N} w_{n}^{(k_n)}.
\end{equation}
Alternatively, we can choose the $k^{th}$ free NN site with the same probability $p_k = 1/l_n$ and replace the ``local'' weight by
\begin{equation}
 w_n^{(k)} = l_{n} e^{-E_k/k_B T}.
\end{equation}
This is the method we use in this work. If $L$ walks are started, then, $Z_N \simeq \sum_{i=1}^{I_N} W(S_i)/L$, where $I_N$ is the number of successfully generated chains of length $N$. Therefore, the expected value of the quantity $O_N$ is $\left\langle O_N \right\rangle = \sum_{i=1}^{L} O(S_i) W(S_i)/\sum_{i=1}^{L} W(S_i)$. 

We notice that, although Rosenbluth method can efficiently sample small chains (typically, $N\lesssim 100$ in the square lattice), for larger ones the attrition of walks becomes a problem. Moreover, the distribution of weights can become so wide that few (rare) chains with large $W_N$ will dominates the statistical averages, leading to unreliable results. 

In the PERM algorithm, the growth proceeds as above, however, at each stage of the growth we make $K$ copies of the chain, if its weight is larger than a parameter $T_n$. Each copy follows growing independently, but, once copies are made, their weights are reduce by a factor $K$. Here we use $K=2$, i. e., we duplicate the chains with $W_n > T_n$. In the same way, at each stage of the growth, we prune chains, with probability $1/2$, whose weight is smaller than a parameter $t_n$. If the chain with $W_n < t_n$ is not pruned, its weight is increased by a factor $2$. As shown in Ref. \cite{Grassberger1}, good results can be obtained fixing $T_n/t_n = 10$, thus, we use this parameter here.

\section{Thermodynamic properties of the model}
\label{results}

We present results for chains with up to 5000 monomers. For each set of NN and NNN energies, we run averages over up to $10^{6}$ started configurations.

The critical properties of the model are obtained in terms of the variables $K_1 \equiv \epsilon_1/k_B T$ and $K_2 \equiv \epsilon_2/k_B T$. Therefore, we will determine the critical ($\Theta$-) points, for example, fixing $K_1$ and varying $K_2$, or vice-versa.

\subsection{Square lattice}

\begin{figure}[!t]
\includegraphics[width=8.cm]{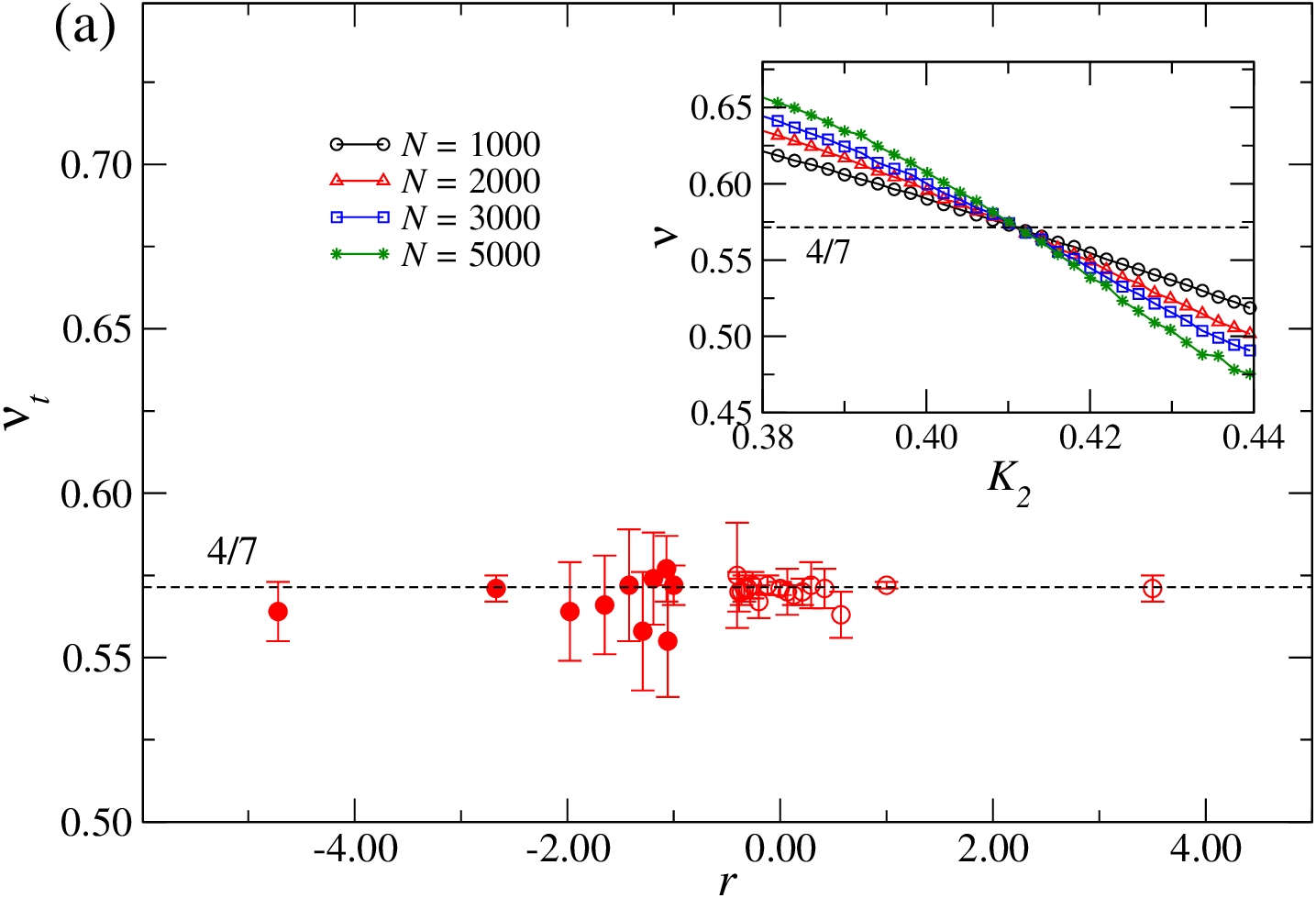}
\includegraphics[width=8.cm]{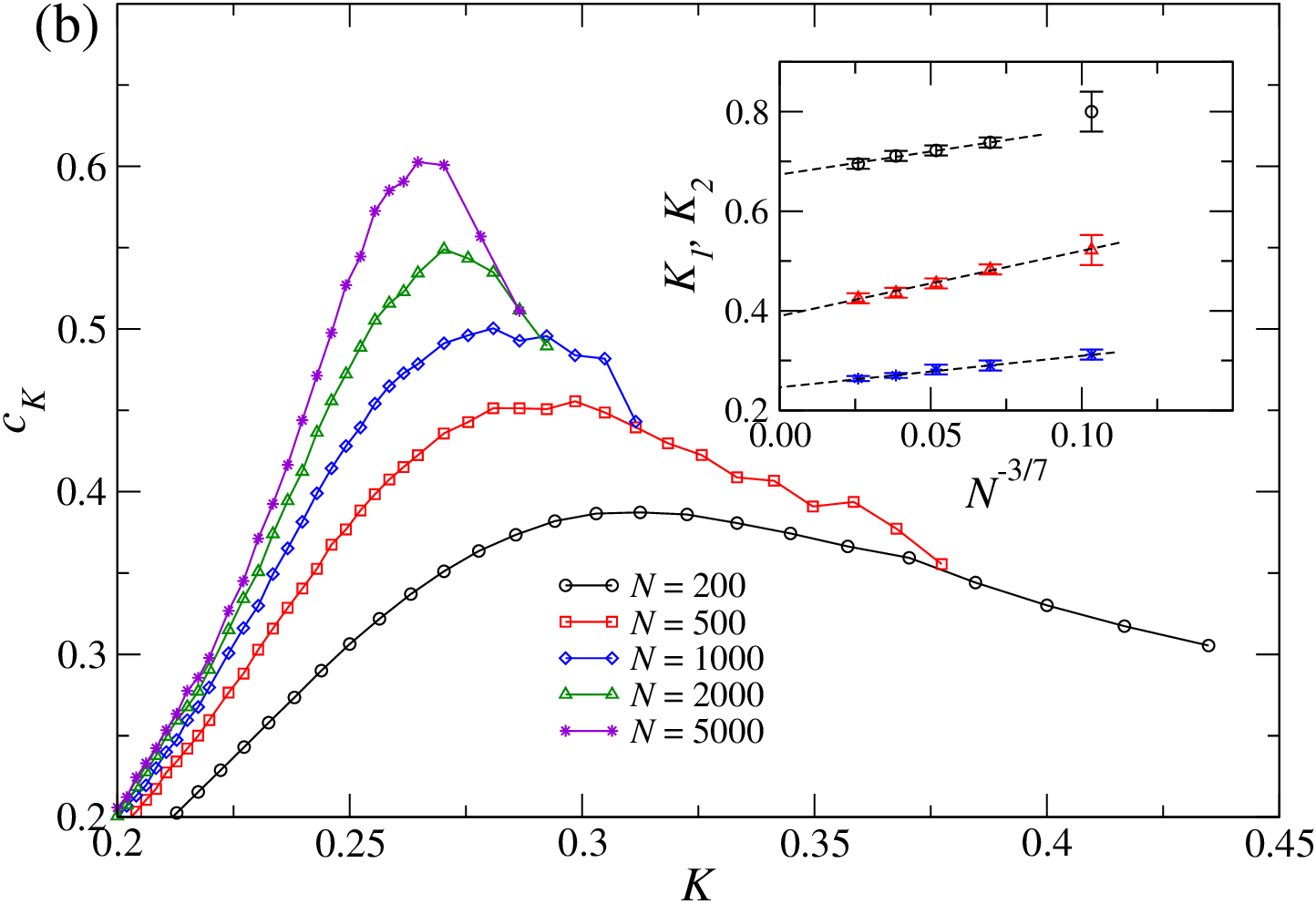}
\includegraphics[width=8.cm]{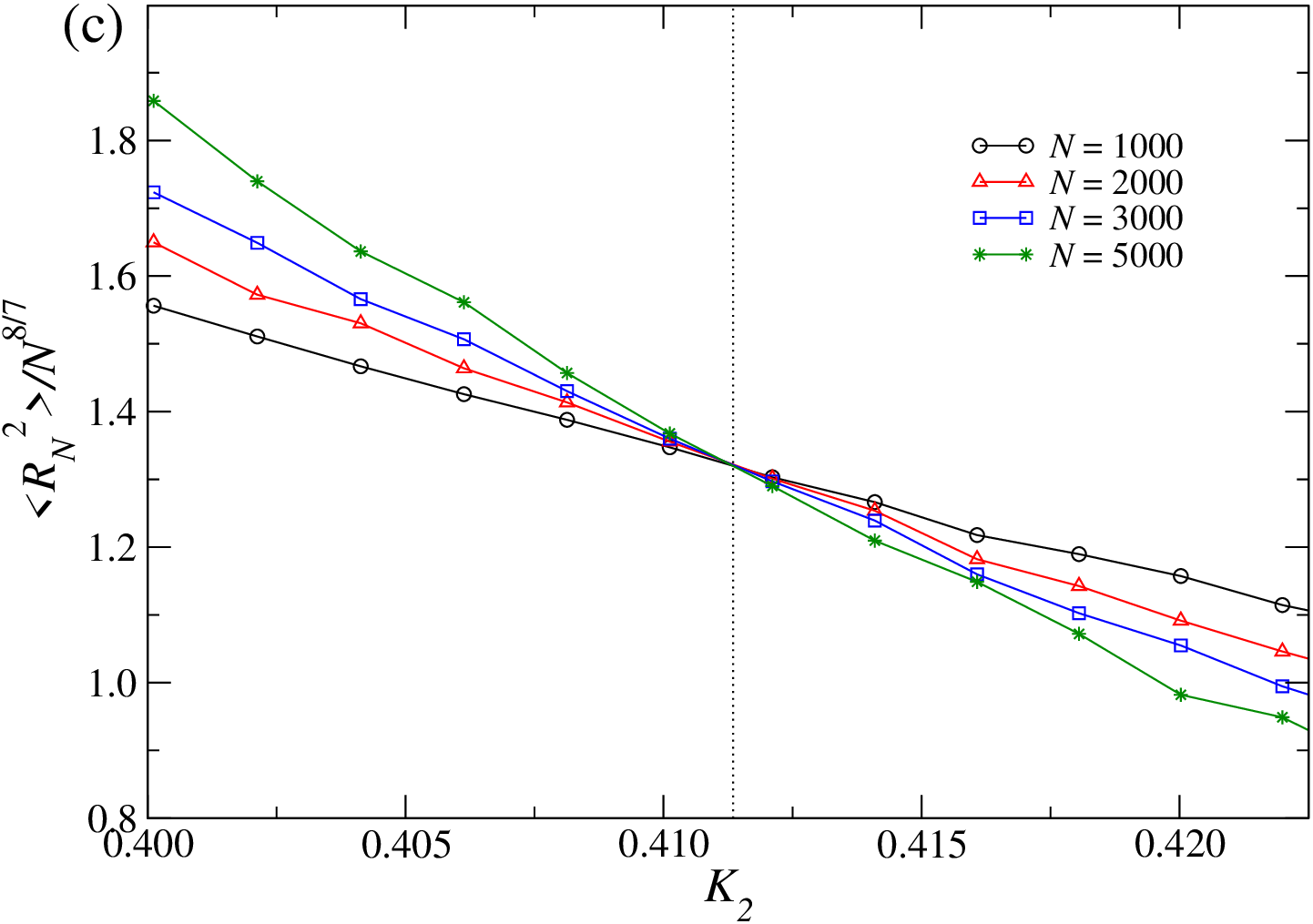}
\caption{(Color on-line) (a) Metric exponents $\nu_t$ against the ratio $r=K_2/K_1$. Open and full symbols indicate the regions of $K_1>0$ and $K_1<0$, respectively. Inset shows the exponents $\nu$ as a function of $K_2$ for $K_1 = 0$ and different polymer lengths. (b) Specific heat per monomer against $K=K_1=K_2$ for different polymer lengths. Inset shows extrapolations of $K_{1,\Theta}(N)$ for $K_2=0$ (black circles), $K_{2,\Theta}(N)$ for $K_1=0$ (red triangles) and $K_{1=2,\Theta}(N)$ for $K_1=K_2$ (blue stars). (c) Rescaled average squared end-to-end distance $\left\langle R_N^2 \right\rangle/N^{8/7}$ against $K_{2}$ for $K_1 = 0$.}
\label{fig1}
\end{figure}

According to Eq. \ref{eqScaling}, from the slope of a log-log plot of the average squared end-to-end distance, $\left\langle R_N^2 \right\rangle$, against the polymer size, $N$, we may estimate the exponent $\nu$. The inset of Fig. \ref{fig1}a shows $\nu(K_2)$ for $K_1=0$ and different chain lengths. Although we observe a dependence of $\nu$ with the polymer length, at critical points it should be independent of $N$. Thus, from the crossing points of the curves for large $N$, we estimate the critical exponent $\nu_{t}$, which is $\nu_t = 0.570(2)$ for $K_1=0$. A similar behavior was found in the whole range of parameters analyzed here $(-1.0 \lesssim K_1 \lesssim 2.0$ and $-0.8 \lesssim K_2 \lesssim 1.0)$, which allows us to determine the exponent $\nu_t$ along the transition line. These exponents are shown in Fig. \ref{fig1}a as a function of the ratio $r=K_2/K_1=\epsilon_2/\epsilon_1$ and are consistent, within the error bars, with the $\Theta$ exponent $\nu_t = 4/7$. This result suggests that the $\Theta$-class is not changed by the NNN interactions, i. e., there exists a $\Theta$-line separating the coil and globule phases. More interestingly, the globule phase as well as the $\Theta$-universality class still exists even for repulsive (NN or NNN) interactions ($r<0$).

\begin{figure}[t]
\includegraphics[width=8.cm]{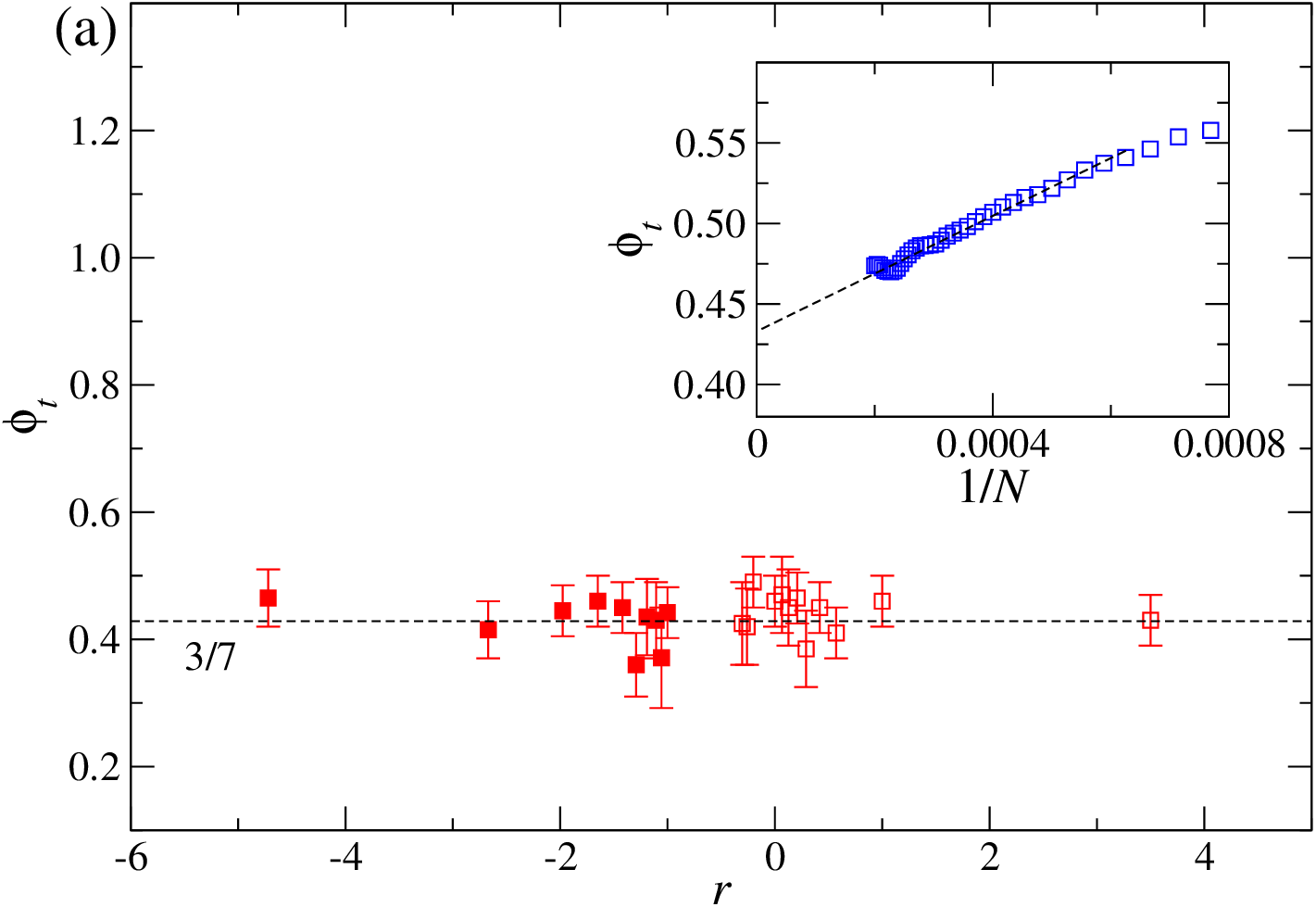}
\includegraphics[width=8.cm]{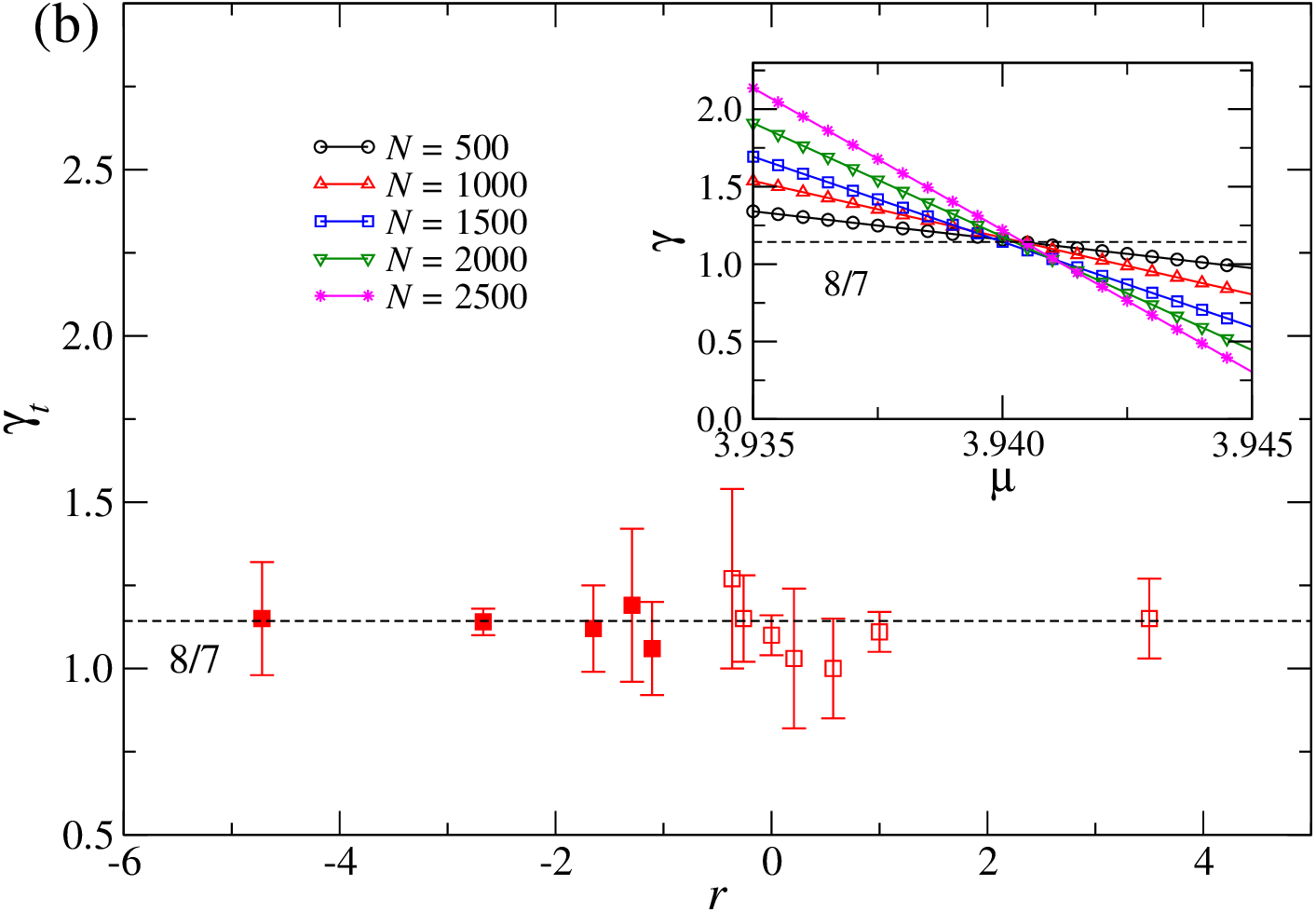}
\includegraphics[width=8.cm]{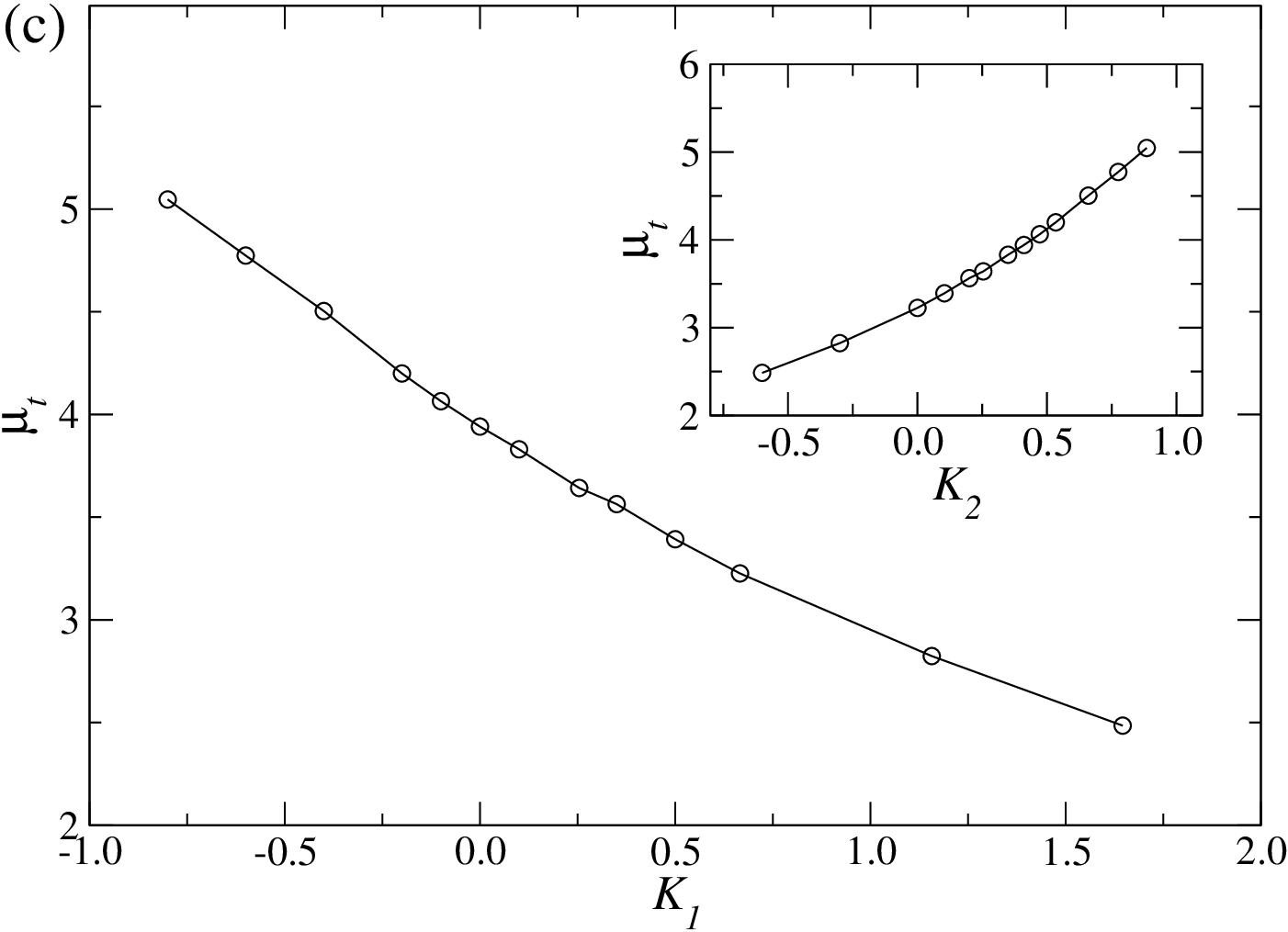}
\caption{(Color on-line) (a) Crossover exponents $\phi_t$ versus the ratio $r=K_2/K_1$. Inset shows the extrapolation of $\phi_t(N)$ for $K_2=0.2$. (b) Entropic exponents $\gamma_t$ against the ratio $r$. The values of $\gamma(\mu)$ for $K_1=0$ are shown in the inset. In (a) and (b), open and full symbols indicate the regions of $K_1>0$ and $K_1<0$, respectively. (c) Growth parameter $\mu_t$ (at the $\Theta$-line) as a function of $K_1$ (main plot) and $K_2$ (inset).}
\label{fig2}
\end{figure}

From the crossing points of Fig. \ref{fig1}a, we estimate $K_{2,\Theta} = 0.411(1)$. In similar plots for $K_1=K_2$ and $K_2=0$, for example, we obtain $K_{1,\Theta} = 0.254(1)$ and $K_{1,\Theta} = 0.666(3)$, respectively. The last value is in good agreement with several previous estimates (see Table I from Ref. \cite{Lee2}), while the other values are slightly larger than the ones reported in Ref. \cite{Lee4}, where the last value is also underestimated, which is possible due to the small polymer lengths considered there ($N \leq 38$). It is interesting to notice that, considering energies and $k_B$ equal unity, the results above leads to $T_{\Theta,\epsilon_2=0} \approx 1.50$, $T_{\Theta,\epsilon_1=0} \approx 2.43$ and $T_{\Theta,\epsilon_1=\epsilon_2} \approx 3.94$. Although the numbers of NN and NNN in the square lattice are both equal to four, attractive NNN interactions will favor the formation of bends in polymer and, hence, the globule phase. Furthermore, this effect will be more pronounced if attractive NN and NNN interactions cooperate to collapse the chain, which explains $T_{\Theta,\epsilon_1=\epsilon_2} > T_{\Theta,\epsilon_1=0} > T_{\Theta,\epsilon_2=0}$.
 
In order to check the reliability of the $\Theta$-points estimated from $\nu(K)$, we calculate the specific heat per monomer $c_{K} = K^{2} (\left\langle E^2\right\rangle - \left\langle E \right\rangle^2)/N$, for the three cases discussed above. For $K_1=K_2(\equiv K)$, this is shown in Fig. \ref{fig1}b, and similar results were found in the other cases. From the peaks of $c_{K}$, we obtain finite-size estimates of the $\Theta$-point, which are expected to approach the asymptotic value as $K_{\Theta}(N) - K_{\Theta}(\infty) \sim N^{-\phi}$. Inset of Fig. \ref{fig1}b shows $K_{1,2,\Theta}(N)$ versus $N^{-\phi}$, with $\phi=\phi_t=3/7$. The good linear behaviors found give indications that $\phi$ is, in fact, the $\Theta$ exponent. Extrapolating these data to $N \rightarrow \infty$, we obtain $K_{\Theta,\epsilon_2=0}=0.67(1)$, $K_{\Theta,\epsilon_1=0}=0.39(3)$ and $K_{\Theta,\epsilon_1=\epsilon_2}=0.25(1)$, in agreement with our estimates above. However, here the error bars are larger, because it is very statistical demanding to obtain smooth $c_{K}$ curves for large polymer lengths.

Equations \ref{eqScaling} and \ref{eqScaling2} state that curves of $\left\langle R_N^2 \right\rangle/N^{2 \nu_t}$ against $K_{1,2}$, for different lengths, will cross at the same point (the $\Theta$-point) if the correct exponent $\nu_t$ is used. Figure \ref{fig1}c presents this quantity as a function of $K_2$, for $K_1=0$, with $\nu_t=4/7$. The almost single crossing points found in these curves give us one more evidence of the $\Theta$ universality class and leads to almost the same values for the $\Theta$-points obtained from $\nu(K)$. Using this method, we obtained the $\Theta$-line.

\begin{figure}[t]
\includegraphics[width=8.cm]{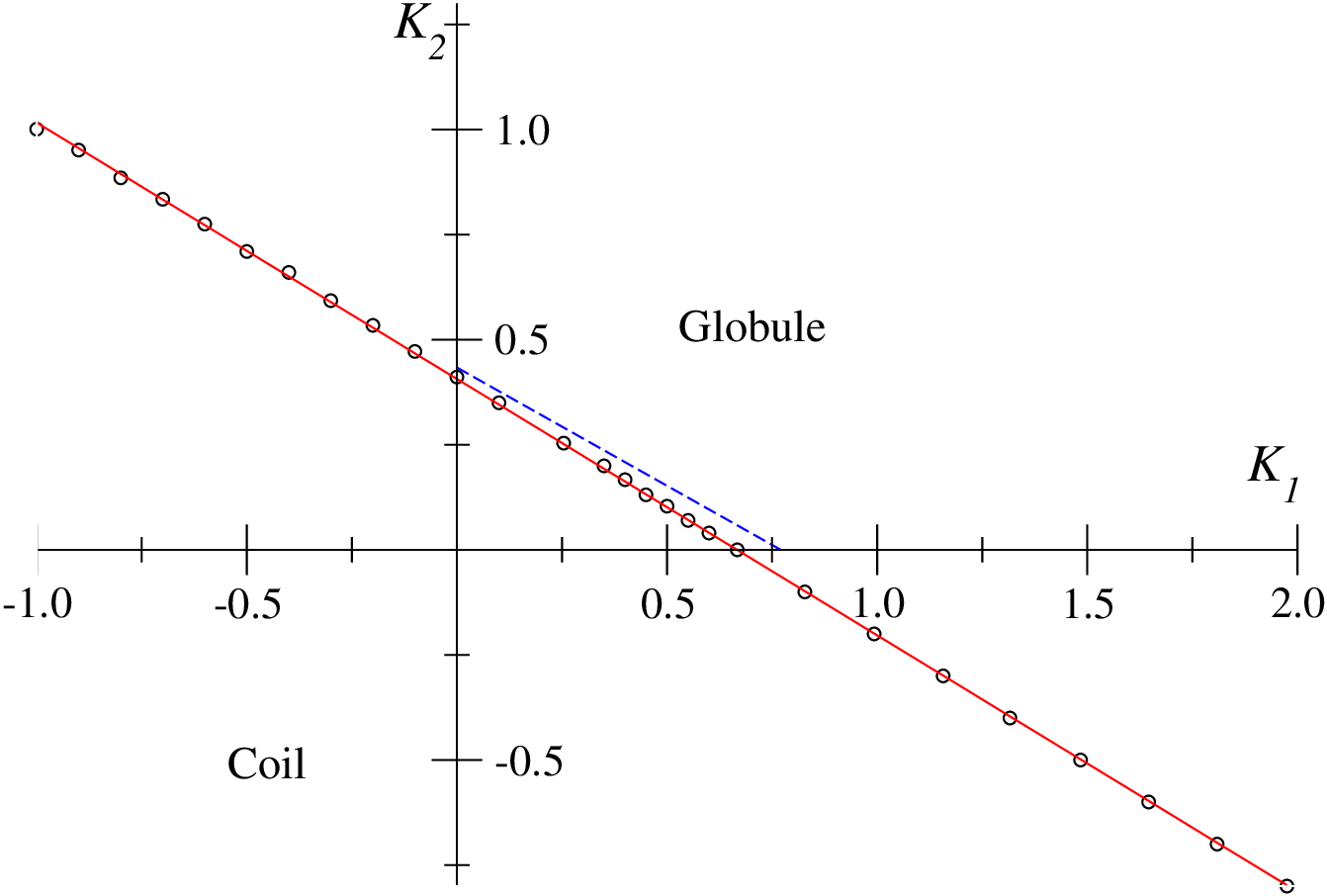}
\caption{(Color on-line) Phase diagram of the generalized ISAW model on the square lattice. The full (red) line is given by Eq. \ref{eqLinha}, while the dashed (blue) line is the result from Ref. \cite{Lee4}.}
\label{fig3}
\end{figure}

From Eq. \ref{eqScaling}, the critical exponents $\phi_t$ can be estimated taking derivatives of $\left\langle R_N^2 \right\rangle$ with respect to $K$ at the $\Theta$-point, which should scale as
\begin{equation}
 R'_N = \dfrac{1}{\left\langle R_N^2 \right\rangle}\dfrac{\partial \left\langle R_N^2 \right\rangle}{\partial K} \sim N^{\phi_t}.
\end{equation}
Hence, from log-log plots of $R'_N$ versus $N$, for different lengths, we estimate $\phi_t(N)$, which is extrapolated to give us $\phi_t$ (see inset of Fig. \ref{fig2}a). Figure \ref{fig2}a shows the exponents $\phi_t$ obtained in this way for different values of the ratio $r$. The good agreement with the $\Theta$ value $\phi_t=3/7$ gives one additional evidence that this model belongs to $\Theta$ universality class.

The partition function of the system is expected to scale as
\begin{equation}
 Z_N \sim \mu^{N} N^{\gamma - 1},
\label{eqFuncPart}
\end{equation}
where $\mu$ is the growth parameter and $\gamma$ is the entropic exponent. This exponent assume the exact value $\gamma_t=8/7$ in the $\Theta$ class and can be obtained from $2 Z_{2 N}/(Z_N \mu^N) = 2^{\gamma}$. Moreover, from the crossing points of curves of $\gamma$ versus $\mu$ for different lengths, we estimate $\gamma_t$ and $\mu_t$ at the $\Theta$-points, as illustrated in the inset of Fig. \ref{fig2}b. The values of $\gamma_t$ as a function of $r$ are shown in Fig. \ref{fig2}b and, again, are in good agreement with the $\Theta$ class. For $K_2=0$, we found $\mu_t = 3.226(1)$, which is very close to the value reported in Ref. \cite{meirovitch}. At variance with the behavior of the tricritical exponents, $\mu_t$ is not constant along the $\Theta$-line. As shown in Fig. \ref{fig2}c, it decreases monotonically with $K_1$ and increases with $K_2$. Interestingly, for $K_1<0$ (where $K_2>0$), we found $\mu_t>4$, i. e., larger than the lattice coordination number.

Finally, the phase diagram of the model is depicted in Fig. \ref{fig3}. The $\Theta$-line has an almost linear behavior, being well-fitted by
\begin{equation}
 K_2 \simeq - 0.6099 K_1 + 0.4066.
\label{eqLinha}
\end{equation}
This behavior is consistent with the findings of Lee et al. \cite{Lee4}, but their line is $K_2 \simeq - 0.563 K_1 + 0.434$, which is slightly above ours (see Fig. \ref{fig3}). Hence, the critical temperatures there are smaller than the ones here, possible due to the small lengths handled in \cite{Lee4}.

\subsection{Cubic lattice}

In the cubic lattice the probability of a polymer chain becoming trapped is smaller than in the square one, thus, it is easier to generate the polymers here. This allows us to analyze a broader range of parameters than in two-dimensions, with $-3.0 \lesssim K_1 \lesssim 2.6$ and $-0.8 \lesssim K_2 \lesssim 0.7$. 

\begin{figure}[b]
\includegraphics[width=8.cm]{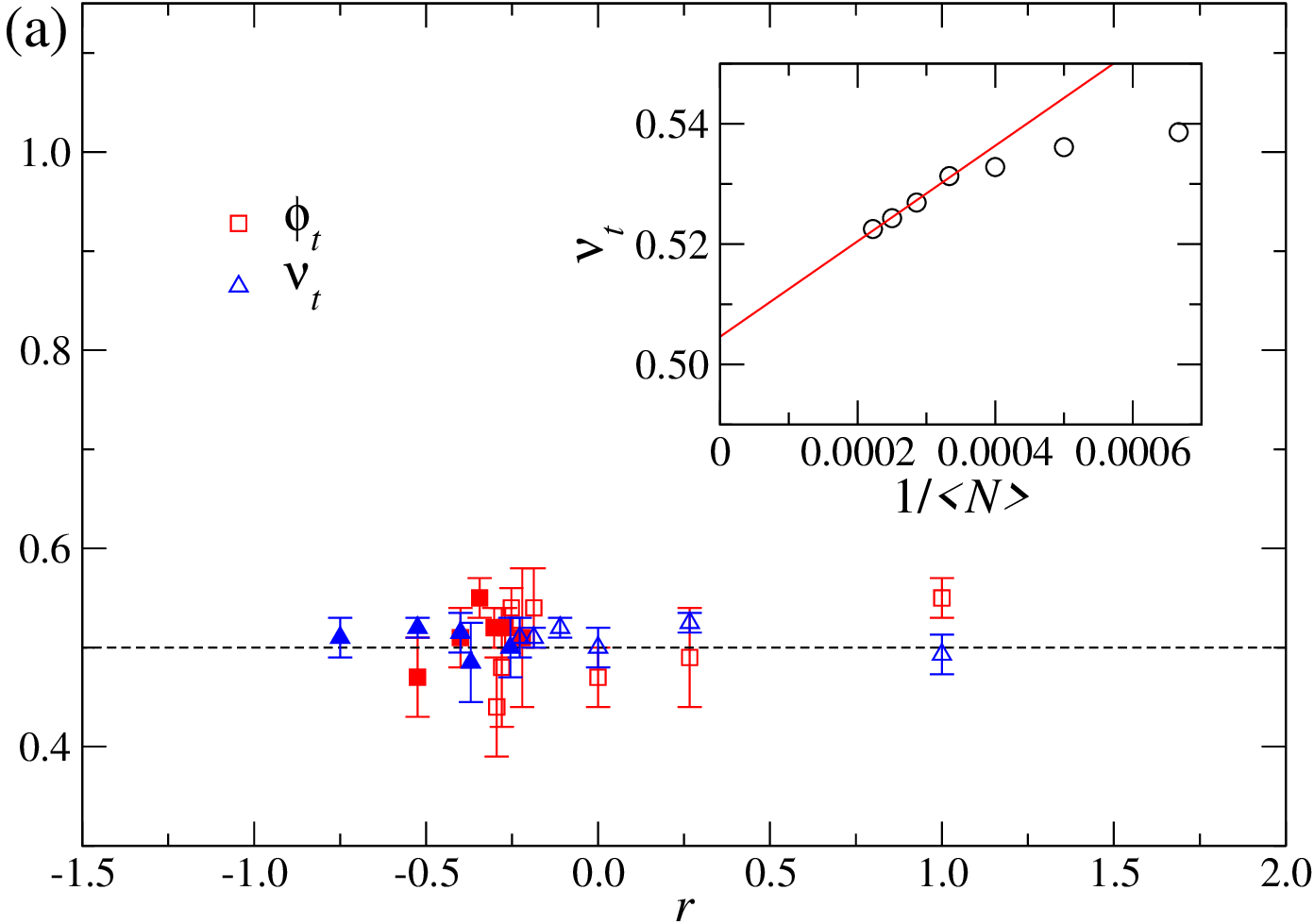}
\includegraphics[width=8.cm]{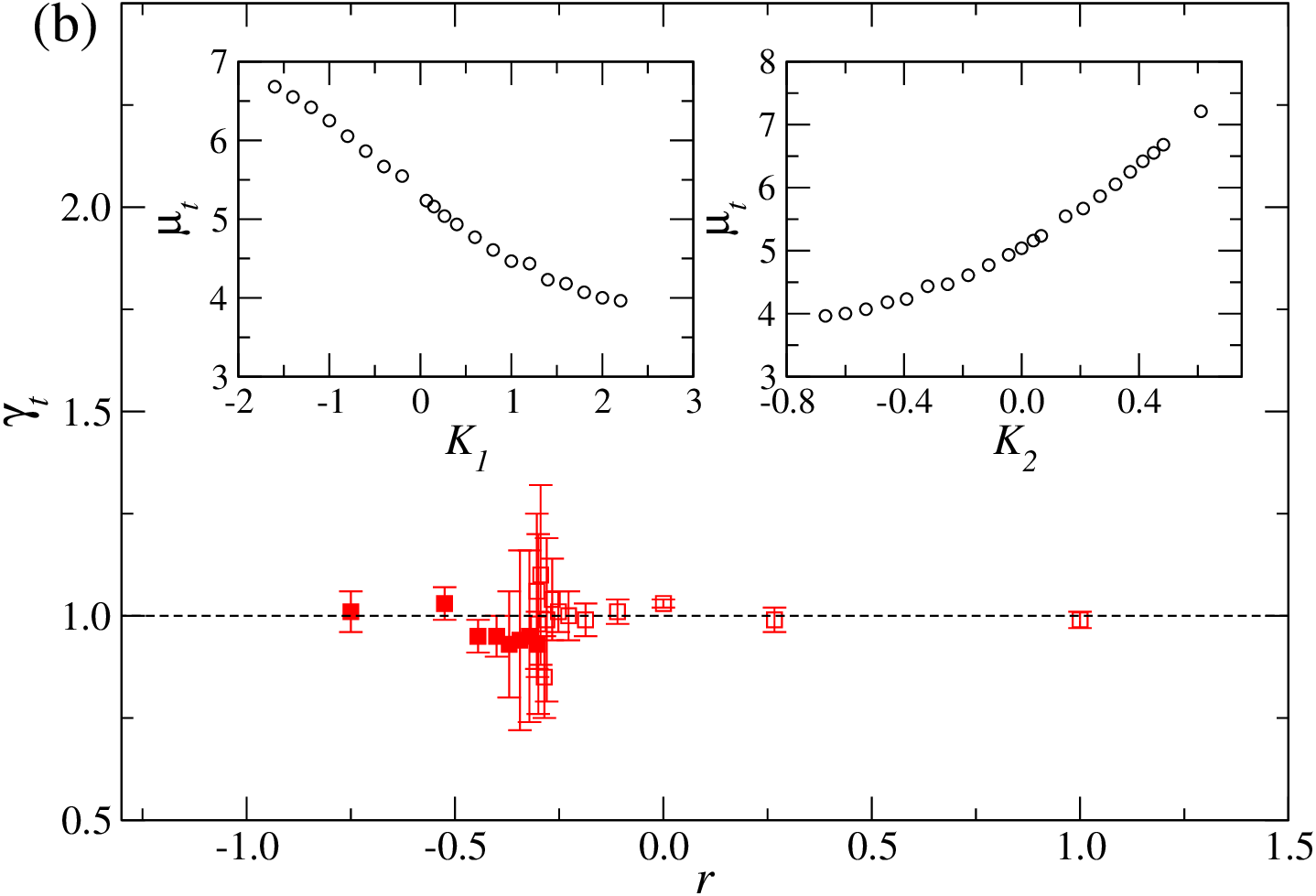}
\caption{(Color on-line) (a) Metric $\nu_t$ and crossover $\phi_t$ exponents versus the ratio $r=K_2/K_1$. Inset shows the extrapolation of $\nu_t(N)$ for $K_2=0$. b) Entropic exponents $\gamma_t$ against the ratio $r$. Insets show the growth parameter $\mu_t$ as functions of $K_1$ (left) and $K_2$ (right).}
\label{fig4}
\end{figure}

Averaging the crossing points of $\nu(K)$ curves for different sizes (similar to Fig. \ref{fig1}a), we obtain metric exponents in the range $0.51 \lesssim \nu_t \lesssim 0.56$, which are slightly larger than the $\Theta$ class one ($\nu_t = 1/2$). However, taking the exponents from the crossing points for pair of sizes ($N_1,N_2$) and extrapolating them to $\left\langle N\right\rangle  \equiv (N_1+N_2)/2 \rightarrow \infty$ (see inset of Fig. \ref{fig4}a), we found exponents in good agreement with the $\Theta$ ones, as shown in Fig. \ref{fig4}a. Following the same procedures of the previous subsection, we determine the critical exponents $\phi_t$ and $\gamma_t$ along the transition line, which are shown in Figs. \ref{fig4}a and \ref{fig4}b, respectively, as functions of the ratio $r = K_2/K_1$. Again, both exponents are in good agreement with the ones of the $\Theta$ class ($\phi_t=1/2$ and $\gamma_t=1$). Thus, as in the square lattice, we conclude that the coil and globule phases are separated by a $\Theta$-line, which exists even for repulsive interactions, showing the robustness of the $\Theta$ class.

\begin{figure}[t]
\includegraphics[width=8.cm]{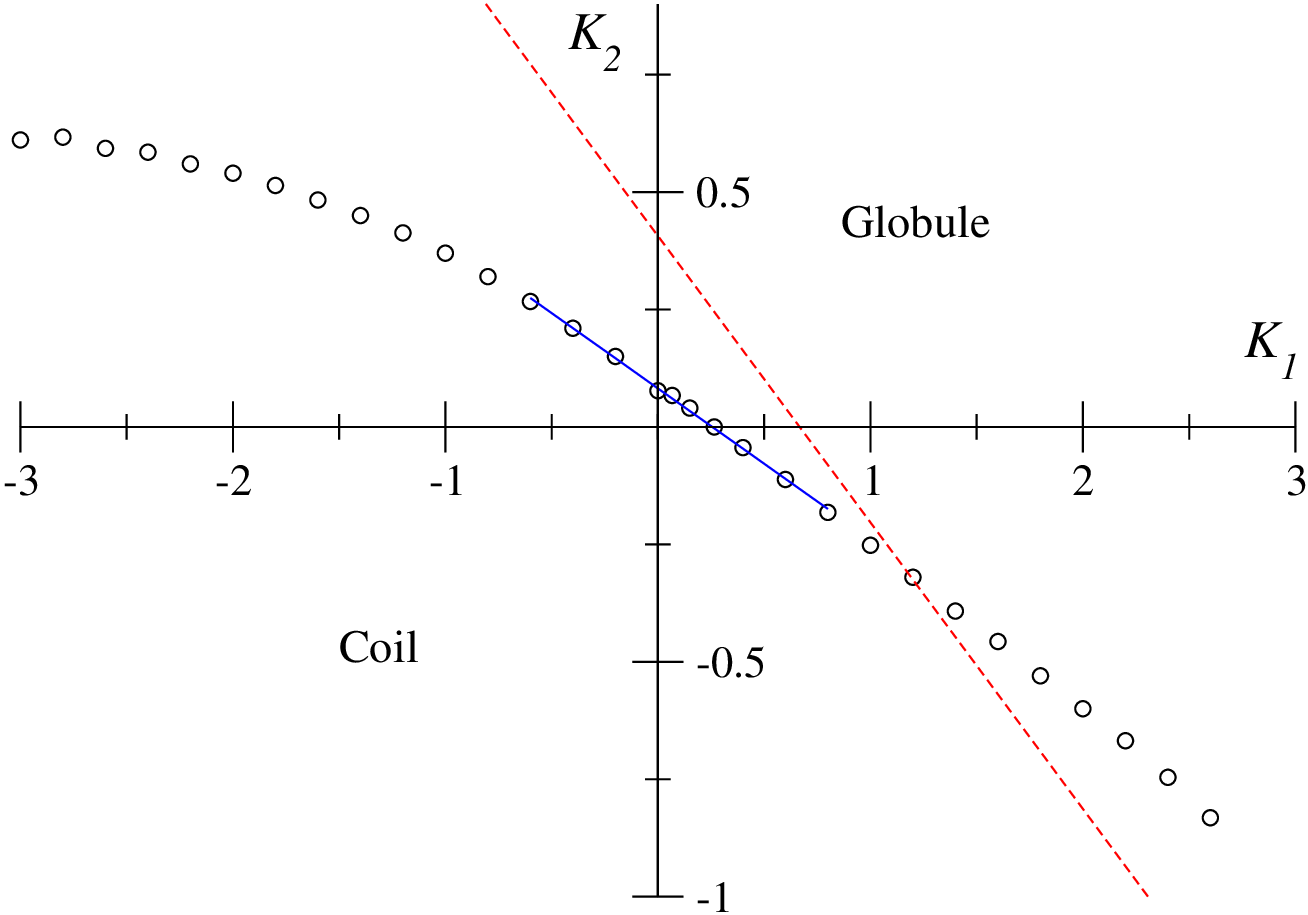}
\caption{(Color on-line) Phase diagram of the generalized ISAW model on the cubic lattice. The dashed (red) line is the $\Theta$-line found in square lattice (Eq. \ref{eqLinha}). The full (blue) line is a fit given $K_2 = - 0.321 K_1 + 0.082$.}
\label{fig5}
\end{figure}

The growth parameters $\mu_t$ along the $\Theta$-line are shown in the insets of Fig. \ref{fig4}b. For $K_2=0$, we found $\mu_t = 5.039(2)$, very close to the value reported in \cite{meirovitch2}. Similarly to the square lattice, $\mu_t$ decreases with $K_1$ and increases with $K_2$, but here there is a clear inflection in $\mu_t$ versus $K_1$ curve. It is noteworthy that in the region of $K_1<0$ it becomes larger than the lattice coordination.

The critical parameters $K_{1,\Theta}$ and $K_{2,\Theta}$, obtained in the same ways of the previous subsection, are summarized in the phase diagram depicted in Fig. \ref{fig5}. In contrast to the square lattice, the whole $\Theta$-line is clearly non-linear here, though an almost linear behavior is observed within and around the attractive region, where $K_2 \approx -K_1/3$ is found. We notice that in the cubic lattice each monomer inside the chain has 4 NN and 12 NNN sites which could be occupied and contribute to the energy of the chain. Thus, in average, for each NN interaction there will be approximately three NNN ones in the globular configuration, explaining $K_1 \approx 3 K_2$ in the $\Theta$-line. The values of $K_{1,\Theta}$ and $K_{2,\Theta}$ are smaller than the ones found in the square lattice, in the region of attractive forces, meaning that the $\Theta$ temperatures are larger here. For example, $T_{\Theta,\epsilon_2 = 0} \approx 3.77$, $T_{\Theta,\epsilon_1 = 0} \approx 12.1$ and $T_{\Theta,\epsilon_1=\epsilon_2} \approx 15.6$, for unitary energies and $k_B$. The first value are in good agreement with the ones typically reported in literature (see Table I of Ref. \cite{Lee1}). The large values of $T_{\Theta}$ in the presence of NNN interactions are expected, since the number of NNN sites is twice the NN ones in the cubic lattice. Thus, the stronger cooperative effects of NNN monomers will facilitate the collapse.

\section{Discussions and conclusions}
\label{conclusions}

In summary, we have studied a generalized ISAW model where next-nearest-neighbor (NNN) interactions exists together the NN ones. Although the NNN energies change the collapse transition points, this transition is found to be always in $\Theta$ class, at least in the set of parameters analyzed. Namely, critical exponents $\nu_t$, $\phi_t$ and $\gamma_t$ in good agreement with the $\Theta$ ones were found along the (transition) $\Theta$-line, in both square and cubic lattices. In the first one, the $\Theta$-line seems to have a linear behavior in the phase diagram, as already observed for this model \cite{Lee4}. Along the $\Theta$-lines, the growth parameters, from the partition function, have monotonic behaviors decreasing (increasing) with $K_1$ ($K_2$).

We remark that a single $\Theta$-line in the whole phase diagrams is \textit{per si} a very interesting result. Although this is expected in the region of attractive forces ($K_1>0$ and $K_2>0$), when attractive and repulsive interactions compete (in the regions $K_1 < 0$ and $K_2>0$ or $K_1 > 0$ and $K_2 < 0$) the existence of a $\Theta$-line is not so obvious. Particularly, in the MMS model (discussed in the Introduction), the competition between attractive and repulsive forces between two and three monomers (in a site) changes the nature of the coil-globule transition in that model \cite{Krawczyk,Tiago}. However, in MMS model the energies are associated to on-site monomers, instead of neighbors as in our model. Moreover, since self-avoidance is relaxed in MMS model there is a large difference between the entropies of both models, which certainly explains the different behaviors found.

As a final remark, we notice that the region with $K_2 < 0$ in our phase diagrams leads to semiflexible polymers, because a repulsive NNN force inhibit the formation of bends in the chains. Therefore, a transition to a crystalline phase could be expected there, which we do not observe. However, the repulsive NNN interaction in our model is different from a bending energy of semiflexible models \cite{Zhou2}, since the first also inhibit the formation of pairs of nonconsecutive NN monomers in simple cubic lattices. Therefore, the change from a $\Theta$-line to a direct (first-order) coil-crystal transition possible does not exists here, or appears in a region of $|\epsilon_1| \gg |\epsilon_2|$, which is very difficult to be accessed with our simulation method.

\acknowledgments

The author thanks JF Stilck for a critical reading of this manuscript and helpful discussions, and the support of CNPq and FAPEMIG (brazilian agencies).

\end{document}